\documentstyle[aps,prc,psfig,epsfig]{revtex}              
\newcommand \be{\begin{eqnarray}} 
\newcommand \ee{\end{eqnarray}} 

\begin{document} 
\draft 
\twocolumn[\hsize\textwidth\columnwidth\hsize 
                    \csname @twocolumnfalse\endcsname  
\title{ Short Range Interaction Effects on the Density of States of Disordered
Two Dimensional Crystals with a half--filled band} 
\author{E. P. Nakhmedov$^{1,2}$  and   V. N. Prigodin$^3$}
\address{$^1$ Max--Planck--Institut f\"ur Physik komplexer
Systeme,N\"othnitzer str.38, 01187 Dresden, Germany\\ 
$^2$ Azerbaijan Academy of Science, Institute of Physics, H. Cavid
str.33, 370141 Baku, Azerbaijan\\  
$^3$  Physics Department, The Ohio State University,
Columbus, OH 43210- 1106, USA} 
\maketitle
\date{\today} 
\maketitle 
\begin{abstract} 
The Density of electronic States (DoS) of a two--dimensional square
lattice with 
substitutional impurities is calculated in the presence of short--range
electron--electron interactions. In the middle of the energy band, the Bragg
reflections off the Brillouin zone boundary are shown to lead to additional
quantum corrections to the DoS, the sign of which is opposite to the
sign of the  
Altshuler--Aronov's logarithmic correction. The resulting quantum
correction to the DoS at half--filling is positive, i.e. the DoS
increases logarithmically as the Fermi energy is approached. However,
far from the commensurate points where the Bragg 
reflections are suppressed, the negative logarithmic corrections to
the DoS survive.  
\end{abstract} 
\vskip2pc] 
 
The competition of the effects of interaction and randomness in
one--dimensional ($1D$) and 
two--dimensional ($2D$) electronic systems is one of the most
intriguing problems  
in low temperature physics. Impurities of arbitrary concentration in a 
$1D$ metal off half--filling have been shown to localise all
electronic states of a non--interacting electron gas,
\cite{berezinskii}. However, the Bragg reflections appearing at
half--filling delocalise the states and enhance the 
density of states (DoS) in the middle of the band, an effect known as 'Dyson
singularity', \cite{lgp,dyson}. A $1D$ correlated electron gas without
disorder is described by the Luttinger liquid theory, \cite{voit}. In
a half--filled band, charge excitations with 
gap (Mott insulator) may be relevant in $1D$ systems as a consequence
of commensurability, 
\cite{schulz},
whereas the localization length for coherent propagation of two
interacting particles in a $1D$ random electronic system  has been
shown \cite{s} to be larger than the one--particle localization
length, i.e., interaction leads to a significant delocalization of the
pair. 

The problem of the interplay of interaction and disorder in $2D$
systems is as complicated as it is in $1D$ 
systems. Indeed, all states of a $2D$ electron gas have been proved to be
localized irrespective of how small the impurity
concentration is, \cite{aalr}. 
Weak Coulomb interactions between
electrons moving diffusively in a $2D$ disordered metal away from
half--filling  have been shown to increase the localization
effect,\cite{aal,aa1,fukuyama}. 
The ground state of a clean $2D$ lattice with nested Fermi surface
becomes unstable with respect
to formation of antiferromagnetic spin gap under an arbitrarily small
Coulomb interaction \cite{hirsch}, and developes a  
Mott gap at strong interactions. Our recent studies of the DoS and of
conductivity of a $2D$
electron gas on a lattice with substitutional
impurities \cite{nko,nfok} have shown
that the commensurability effects at half--filling for the noninteracting
case are opposite to those obtained for $1D$ systems,\cite{wc,gd,gm}, so
that the impurity scatterings with coherent Bragg reflections have
been found to lower the electronic density of states around the Fermi
level. Interaction effects in commensurate weakly disordered $2D$
electronic systems have not yet been studied properly. Notice that,
the recently observed metal--insulator transition (MIT) in $2D$
electronic systems,which 
occurs at low temperatures ($\sim T \le 2 K $),\cite{aks,kkf}, cannot
be explained in the framework of the 'conventional' localization
theory 
and still is one of the 
puzzling issues of central importance in the physics of disordered
systems. Measurements of 
resistivity in high--mobility Si--MOSFET's show that the insulator behavior at
low particle densities, $n < n_c$, crosses over to the metallic one as
the particle density $n$ reaches the critical value $n_c = 9,02\times 10^{10}
cm^{-2}$ ( for $n > n_c$). The observation of the MIT in
$GaAs/AlGaAs$, \cite{rjh}, where e-e interactions are estimated to be
weak, shows that the effect of correlations, although crucial, is not
the only factor leading to MIT. 
 
In this Letter we study the effect of short--range repulsive interactions
on the one--particle DoS of a 2D square lattice with
substitutional impurities for half--filled energy band. We will show
that a class of quantum corrections to the DoS, negative far from
half--filling, changes its sign as the center of the band is
approached. Such a behavior of the DoS is similar to that observed for
the conductivity $\sigma(T)$ \cite{aks,kkf,rjh}. A recent computation
of the conductivity in the half--filled Hubbard model with disorder
also displays a change in the sign of $d\sigma/dT$ as the system
acquires particle--hole symmetry,\cite{dst}.  

The Hamiltonian of interacting electrons in the random field of
substitutional impurities can be written in the Bloch--state
representation in the following form, 
\begin{eqnarray}
\hat{\bf H}&=&\sum_{{\bf p},\sigma} \epsilon ({\bf
p})c_{{\bf p},\sigma}^{+}c_{{\bf p},\sigma} +\nonumber\\
&+& \sum_{{\bf p},{\bf q},{\bf
G},\sigma } \rho_{imp}({\bf q})V_{imp}({\bf q}+{\bf G})c_{{\bf
p},\sigma}^{+}c_{{\bf 
p}+{\bf q}+{\bf G}, \sigma}+ \nonumber\\ 
&+& \sum_{{\bf k},{\bf k'},{\bf q}, {\bf G}, \sigma ,{\sigma}'} U({\bf q}
+{\bf G})c_{{\bf k},\sigma}^{+}c_{{\bf k'},{\sigma}'}^{+}c_{{\bf k'}-{\bf q},
{\sigma}'}c_{{\bf k}+{\bf q}+{\bf G}, \sigma},
\label{ham}
\end{eqnarray}
where
\begin{equation} 
\epsilon ({\bf p})= t[ 2 - \cos(p_{x}a) - \cos(p_{y}a)],
\label{es}
\end{equation}
with $t$ and $a$ being the tunnelling integral for nearest--neighbor sites and
the lattice spacing, respectively.
$V_{imp}({\bf q})$ and $U({\bf q})$ in Eq.(\ref{ham}) are the
Fourier--transforms of a 
single impurity potential and of the short--range e-e interaction potential
respectively; ${\bf G}$ is a reciprocal lattice- vector.
${\rho}_{imp}({\bf q}) = L^{-2}\sum_{\alpha}exp(i{\bf q}{\bf R_{\alpha}})$,
where $L$ is a linear dimension of the system, and ${\bf R}_{\alpha}$
is the coordinate of an impurity, randomly located on a lattice site.
The impurity concentration is assumed to be small, and scatterings on the
$\delta$--correlated impurity potential can be estimated in the framework of
the Born approximation,\cite{agd}. For a metallic system $p_F l \gg 1$
and crossed impurity lines of higher order in $(p_F l)^{-1}$ are
ignored; $p_F$ and $l$ are the Fermi momentum and mean free path,
respectively. 

By varying the band filling, Bragg reflections of the electronic wave off the
Brillouin zone boundary are intensified for commensurate values of the 
electron wavelength,$\lambda$, and the lattice constant,$a$. We will study the
effects of Bragg reflections at half--filling, since they become essential as
the middle of the band is approached. The Fermi surface for the energy
dispersion given by Eq.(\ref{es}) is flat at half--filling, and its
whole section is nested, with  vectors ${\bf Q} = \{\pm \frac{\pi}{a},
\frac{\pi}{a}\}$.
The perfect nesting of the Fermi surface gives rise to the following electron-
hole symmetry relation for the electron dispersion with respect to ${\bf Q}$:
\begin{equation}
\epsilon({\bf p} + {\bf Q}) - {\epsilon}_F = - [\epsilon({\bf p}) -
{\epsilon}_F], 
\label{e-h}
\end{equation}
where ${\epsilon}_F$ is the Fermi energy and ${\epsilon}_F = 2t$ for
the half--filling case. 
 Notice that introducing the next--nearest--neighbor  hopping term
with amplitude $t'$ destroys the perfect nesting of the Fermi
surface. 
However, an optimal nesting takes place with ${\bf Q^*} = 0.91{\bf Q}$
for $t'/t=0.165$, \cite{rrtt}. 

For a small band filling and far from the rational points, the Fermi surface
of the model we are studying looks like a sphere, so that the effects
of periodicity can be 
incorporated into the electronic effective mass. This situation does
not differ from 
that of an electron gas moving in the random field of impurities. The
quantum corrections to the DoS for this model 
have been calculated in \cite{aal,aa1,fukuyama}. The lowest order
quantum correction to the DoS resulting from the  
interactions in the diffusion channel can be expressed in the
following form,\cite{aal,aa1}: 
\begin{eqnarray}
\delta \rho_{N}^{(D)}(\epsilon , T)&=& -\rho_o^{(2d)}\frac{[U(0,0) -
2\overline{U({\bf p}-{\bf p'},0)}]}{4{\pi}^2\hbar D}\nonumber\\
&&\times \ln \frac{1}{\tau_o
max\{\epsilon , T\}},
\label{corAA}
\end{eqnarray} 
where $D$ is the diffusion coefficient, $D =
\frac{{v_F}^2{\tau}_o}{2}$, with $v_F$ and ${\tau}_o$ being the 
Fermi velocity and the impurity relaxation time for Normal
scattering, respectively, and  
${\rho}_0^{(2d)} = \frac{2}{(\pi a)^2 t}\ln ({\epsilon}_F min\{\tau_o,
\frac{1}{|\epsilon |}\})$ .
The bar over the interaction potential $U$ denotes the average over the Fermi
surface, and the additional prefactor $2$ in the Hartree correction term comes
from the spin degeneracy.
Interactions in the Cooper channel give the following contribution to the
DoS far from half--filling,
\begin{equation}
\delta \rho_{N}^{(C)}(\epsilon ,T)= - \frac{1}{2{\pi}^2 \hbar D}\ln \frac{\ln
  T_c\tau_o / \hbar}{\ln T_c/ max\{ \epsilon ,T\}},
\label{rhoNC}
\end{equation}
where $T_c = \epsilon_F \exp (1/\lambda)$ and $\lambda = \rho_o^{(2d)}U$
is the dimensionless interaction constant. Notice that
the dynamical screening  of a short--range interaction in the
diffusion channel far from half--filling does not change the bare
interaction potential considerably, whereas a short range potential is
strongly renormalized in the Cooper channel, \cite{agd,larkin} as seen
performing the   
summation of a ladder series in the bare interaction, which replaces
the logarithmic energy or temperature 
dependence of the quantum correction by the rather weak double
logarithmic dependence given by Eq.(\ref{rhoNC}).  

A new kind of quantum corrections to the DoS takes place in the
correlated disordered system  due to the Bragg reflection, which
enhances the number of electronic states at half--filling. 

As it is well known, the quantum interference corrections to the
thermodynamic and kinetic coefficients come from the singular impurity
ladder series referred to as the \emph{diffuson} and the
\emph{Cooperon} blocks,\cite{aa1}.The diffuson (Cooperon) block has
the diffusion pole when the difference ${\bf q}$ of the momenta of the
electron and the hole (total sum ${\bf k}$ of the momenta of the
electrons) and their energies difference $|\omega_m|$ are small,
i.e. the block acquires the pole in the  diffusion regime when $ql \ll
1$ ($ kl \ll 1$) and $|\omega_m| \tau_o \ll 1$.   

New singular impurity blocks take place at half--filling with
particle--hole symmetry, which are referred to as $\pi$--Diffuson and
$\pi$--Cooperon, \cite{nko,nfok}. The $\pi$--Diffuson
($\pi$--Cooperon) has a diffusion pole at large $\propto {\bf Q}$
momenta differences (total momenta) and small total energies of the
electron and the hole (of two electrons). 
The diagram equations for the $\pi$--Cooperon, $C_{\pi}({\bf
q},i{\omega}_m)$, and for the  
$\pi$--Diffuson, $D_{\pi}({\bf q},i{\omega}_m)$, are given in
Fig.1e,f.
The bare Green's function $G_0$ is indicated in the diagrams by a
straight line, 
and $G_0({\bf p}, i{\epsilon}_n)=\frac{1}{i{\epsilon}_n - (\epsilon ({\bf
p}) - {\epsilon}_F)+ \frac{i}{2{\tau}_o}sign \epsilon_n}$ .
The Green's function of an electron with large momentum is
represented by a dashed line. Notice that the dashing of the line has a
physical meaning only for an impurity vertex or an interaction one when a large
momentum transfer ($\propto {\bf Q}$ due to the Bragg reflection) is
involved in the 
scattering process,\cite{nko,nfok}. Therefore, the new selection of
diagrams to be included in the summation is 
performed according to the rule that a straight line in each impurity
or interaction 
point joins a dashed line and vice versa. This rule is consistent with the fact
that each scattering with large momentum transfer implies also
coherent Bragg reflection of the 
scattered electron on the
boundary of the Brillouin zone. It is easy to see that the momentum
conservation for impurity vertices with large momentum transfer is  violated,
so that they correspond to Umklapp scattering. As far as scatterings on point--
like impurities are considered, these vertices are also characterised
by $\tau_o$. 
By summing the ladder series in Fig.1f, treating the $\pi$--scattering
of an electron on 
impurities perturbatively, the following expression for
the $\pi$--Cooperon $C_{\pi}({\bf q}, i{\omega}_m )$ is
obtained:
\begin{eqnarray}
C_{\pi}({\bf q}, i{\omega}_m )&=&\frac{1}{2\pi
\tau_o \rho_o^{2d}} \bigg\{ \theta (- {\epsilon}_n ({\omega}_m -
{\epsilon}_n )) +\nonumber\\ 
&+& \frac{\theta (\epsilon_n (\omega_m -
\epsilon_n ))}{(1 + |{\omega}_m | {\tau}_o )^2 + (ql)^2 -
1} \bigg\}
\label{pic}
\end{eqnarray} 
The expression for the $\pi$--Diffuson  $D_{\pi}({\bf q},
i{\omega}_m)$ is also given by Eq.(\ref{pic}) with the exception that ${\bf q}$
in the equation for $D_{\pi}({\bf q}, i{\omega}_m )$ denotes
the differences of the momenta of a particle-hole pair.
\begin{figure}
\begin{center}
\epsfxsize88mm \epsfbox{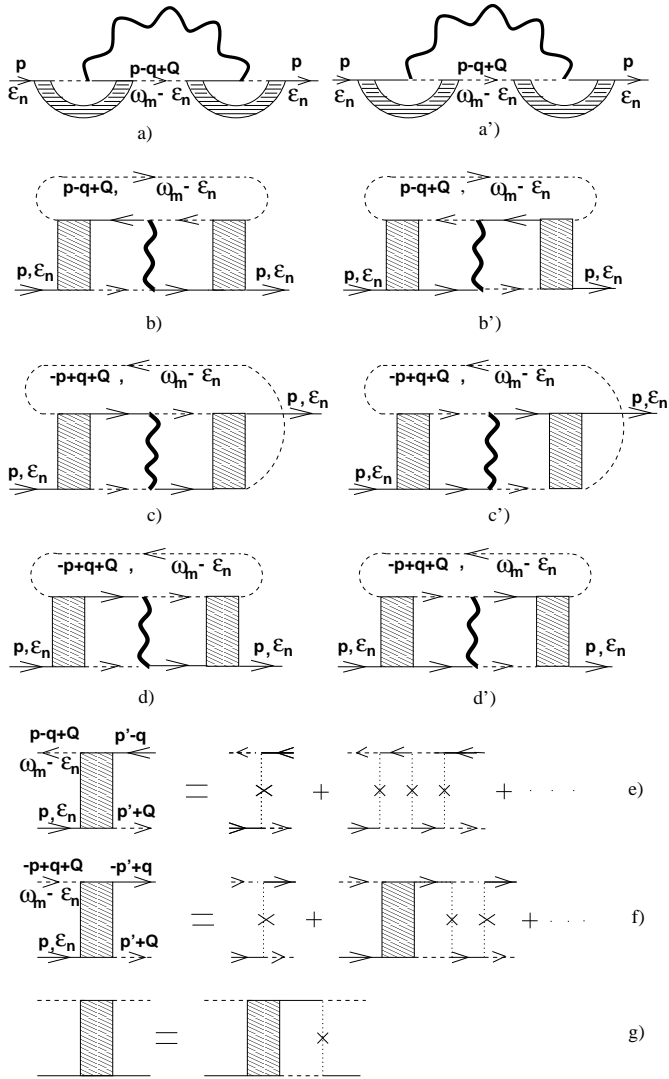}
\end{center}
\caption{Bragg reflection induced contributions to the DoS at
half--filling from $(a),(a');(c),(c')$ exchange and
$(b),(b');(d),(d')$ direct interactions. 
The diagrams obtained from the exchange ones by interchanging the straight
and dashed lines joined in one interaction vertex with appropriate
screening of the Coulomb interaction in particle - hole channel also give
a contribution to the DOS.
(e) and (f) are the impurity ladder series for the 
$\pi$-Diffuson,$D_{\pi}({\bf q},i\omega_m)$, and $\pi$- Cooperon,
$C_{\pi}({\bf q}, i\omega_m)$, respectively. (g) Diagram equation for the block
$\widetilde{D}_{\pi}$ (or $\widetilde{C}_{\pi}$), obtained by adding
one impurity line to the $\pi$-Diffuson (or $\pi$- Cooperon). Heavy
wave line and dotted line with cross denote Coulomb potential and
impurity scattering, respectively.}
\label{fig1}
\end{figure} 
The additional contributions to the DoS in the middle of the band come
from the diagrams shown in Fig.1a-d'. 
The correction to the DoS, $\delta\rho (\epsilon,T)$, due to the e-e
interactions is given by the following 
expression,
\begin{equation}
\delta\rho (\epsilon,T)= -\frac{2}{\pi}\big \{Im \int
\frac{d^2p}{(2\pi)^2}G_o^2({\bf 
p},i{\epsilon}_n)\Sigma_{ee}({\bf p},i\epsilon_n)\big \}_{i\epsilon_n
\to \epsilon}
\label{dos}
\end{equation}
where  $\Sigma_{ee}$ is the self--energy for first order quantum
corrections to the DoS given by Figs.1($a$)-($d'$).
The expression corresponding to the diagram Fig.1a
can be reduced to the following form after simple calculations,
\begin{eqnarray}
\delta {\rho}_a(\epsilon, T)&=& 2
\tau_o^2{\rho}_o^{2d}Im \int
\frac{d^2q}{(2{\pi})^2} \int_{0}^{\infty} \frac{d \omega}{2\pi} U({\bf
  q}-2{\bf Q},2\epsilon -\omega)\nonumber\\
&& \times \frac{\tanh {\frac{\omega + \epsilon}{2T}} + \tanh{\frac{\omega -
\epsilon}{2T}}}{\big [ (1 - i\omega {\tau}_o)^2 + (ql)^2 -1\big]^2} .
\label{cora2}
\end{eqnarray} 
The bare DoS $\rho_o^{(2d)}$ which is involved in the new correction
(8) at half--filling is again given by $\rho_o^{(2d)}=\frac{2}{(\pi
a)^2t}\ln(\epsilon_F min\{\tau_o,1/|\epsilon |\})$, where the van Hove
singularity is cut--off due to the $i/2\tau_o$ term in the bare
Green's function, \cite{comment}. 
Calculations of other diagrams in Fig.1 are similar to the calculation
of $ \delta {\rho}_a(\epsilon, T)$. Also, the contributions from the
diagrams $a')$, $b')$, 
$c')$ and $d')$ in Fig.1 are equal to those coming from $a)$, $b)$,
$c)$ and $d)$, respectively. Screening of the potential $U$ in the
diffusion and cooperon channels is realized according to the
diagrammatic equations given in Fig.2.
\begin{figure}
\begin{center}
\epsfxsize88mm \epsfbox{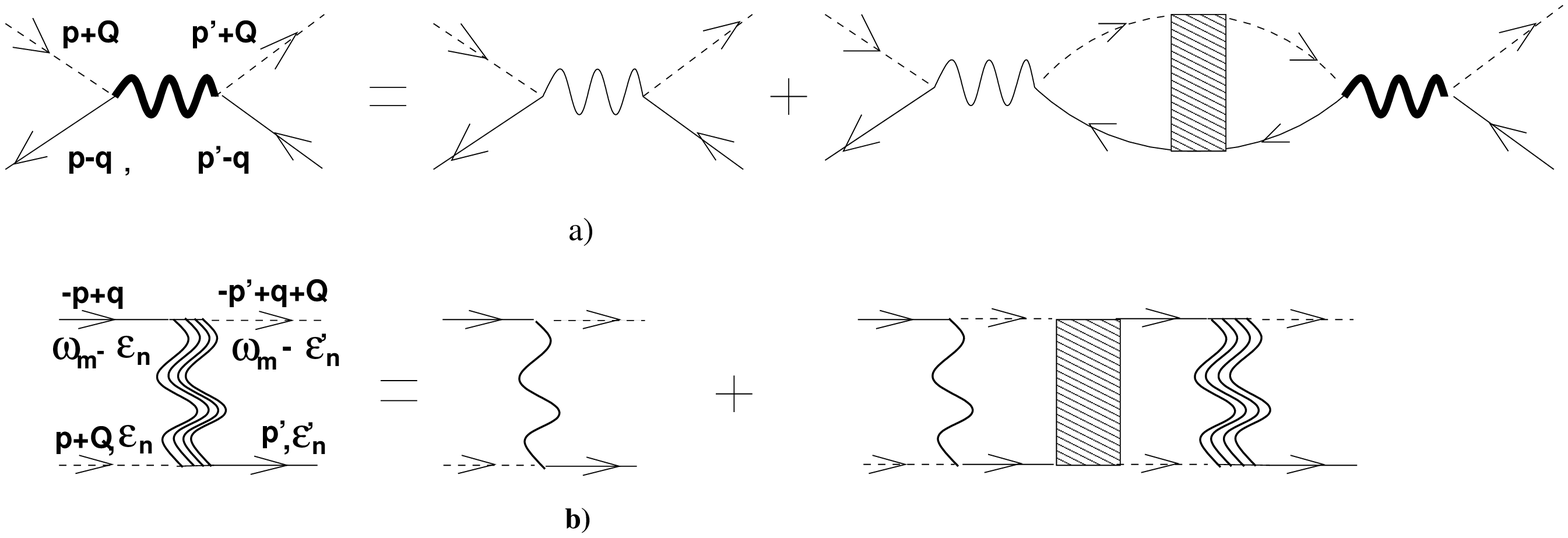}
\end{center}
\caption{Screening of the Coulomb interactions for electrons on a $2D$
  half--filled square lattice in (a) particle--hole  and
  (b)particle--particle channels, respectively.}
\label{fig.2}
\end{figure}
In the middle of the band, it can be shown that the screening of a
short--range interaction arises in the diffusion channel, instead of
the screening of the potential in the  Cooper channel 
which takes place in the electron gas model,\cite{agd,larkin}. It
follows that a dynamical screening of the interaction in the diagrams
$(c)-(d')$ of Fig.1 
does not essentially change the value of the bare potential, and the
contribution to the DoS from these diagrams is calculated according to
Eq.(8),
\begin{eqnarray}
\delta {\rho}_{\pi}^{(C)}(\epsilon, T)&=& {\rho}_o^{(2d)} \frac{[
U(2{\bf Q}, 0) + 
2\overline{U({\bf p} - {\bf p'} +2{\bf Q}, 0)}]}{4{\pi}^2 \hbar D}\nonumber\\
&& \times \ln\big(\frac{1}{{\tau}_o max \{ \epsilon ,T \}}\big) 
\label{corpiC}
\end{eqnarray}
The screening of the interaction in the diffusion channel is realized
according to the diagrammatic equation in Fig.2, 
\begin{equation}
U({\bf q-2Q},\omega_m) = \frac{2}{\rho_o^{(2d)}\big[\ln \frac{T_c}{T}
- \psi(\frac{|\omega_m| +D q^2}{4\pi T} +\frac{1}{2}) +
\psi(\frac{1}{2})\big]}  
\label{scr}
\end{equation}
The correction to the DoS from the diagrams in Fig.1($a,a'$) is
calculated by putting Eq.(\ref{scr}) into Eq.(\ref{cora2}). The
contribution from these diagrams is given   
by the corresponding expression for
$\delta \rho_{N}^{(C)}(\epsilon,T)$, expressed by Eq.(\ref{rhoNC}). The
potential of the other diagrams in the diffusion channel, given by
Fig.1($b,b'$), 
carries zero energy and large momentum $\sim {\bf p'-p'' +2Q}$. 
Therefore screening is not  effective and these diagrams give
logarithmic contribution to the DoS. Therefore, the  
total quantum correction to the DoS in the diffusion channel can be
presented as:
\begin{eqnarray}
&&\delta \rho_{\pi}^{(D)}(\epsilon ,T) = - \frac{1}{2\pi^2 \hbar D}\ln{\frac
{\ln T_c\tau_o/\hbar}{\ln T_c/max\{\epsilon, T\}}} +\nonumber\\
&+& {\rho}_o^{(2d)}
  \frac{\overline{U({\bf p} - {\bf p'} +2{\bf Q}, 0)}}{2{\pi}^2 \hbar D}
\ln\big(\frac{1}{{\tau}_o max \{ \epsilon ,T \}}\big) 
\label{corpiD}
\end{eqnarray}
It is seen from Eqs.(\ref{corpiC}) and (\ref{corpiD}) that the Bragg
reflection  contribution to the DoS
$\delta {\rho}_{\pi}=\delta {\rho}_{\pi}^{(C)} +\delta {\rho}_{\pi}^{(D)}$
increases as $\epsilon$ approaches the Fermi level. 

The total quantum correction to the DoS
$\delta {\rho}(\epsilon, T)$ in the middle of the band can be expressed as
$\delta {\rho}(\epsilon, T) = \delta {\rho}_N(\epsilon, T) + \delta
{\rho}_{\pi}(\epsilon, T)$. Both contributions to the DoS have the same
energy  or temperature dependences. However, they differ in sign and by the
values of the interaction potentials $U(0,0)$ and $U(2{\bf Q},0)$
. Since the short--range interaction is a  
screened Coulomb interaction, the potentials $U(0,0)$ and $U(2{\bf Q},0)$ being
included in Eq.(\ref{corAA}) and Eq.(\ref{corpiC}) respectively cannot
differ strongly from each 
other if the Thomas--Fermi screening number is of the order of the reciprocal
lattice vector $2{\bf Q}$. As a result, the total quantum corrections
to the DoS become positive at half--filling. Far from half--filling
the Bragg reflections are destroyed and the 
dominant contribution to the DoS is the logarithmically decreasing
Altshuler--Aronov contribution \cite{aal,aa1} given by Eq.(\ref{corAA}).  
Notice that an anomaly in the one--particle DoS is experimentally
observable by  measuring  the tunneling conductivity in a contact as a
function of the bias voltage. This dependence completely reflects an
energy dependence of the DoS.  

In conclusion, we showed that the DoS of a $2D$ disordered lattice with
flat Fermi surface is enhanced close to half--filling due to quantum
corrections mediated by  
Bragg reflections. Since similar quantum
corrections exist for all commensurate points, the DoS seems to show
oscillating behavior with maximum amplitude at half--filling. It is
appropriate to emphasize  that, although the Bragg reflection results in the 
decreasing the DoS of a non--interacting system, \cite{nko,nfok}, it gives rise
to an increase  of the DoS in the presence of $e-e$ correlations.  

The work was partially supported by ONR under Grant N00014-01-1-0427.
 
\end{document}